\documentclass{aastex62}

\received{\today}
\revised{}
\accepted{}
\submitjournal{ApJL}
\shorttitle{Exact Calculation of Nonideal Fields}
\shortauthors{Totorica et al.}

\usepackage{graphicx} \usepackage{amsmath} \usepackage{epsfig}

\begin{document}

\title{Exact Calculation of Nonideal Fields Demonstrates Dominance of Injection in Relativistic Reconnection}

\correspondingauthor{Samuel R. Totorica}
\email{totorica@princeton.edu}

\author{Samuel R. Totorica}
\affiliation{Department of Astrophysical Sciences, Princeton University, Princeton, New Jersey 08544, USA}
\affiliation{Department of Astro-fusion Plasma Physics (AFP), Headquarters for Co-Creation Strategy, National Institutes of Natural Sciences, Tokyo 105-0001, Japan}
\affiliation{Princeton Center for Heliophysics, Princeton, New Jersey 08543, USA}
\author{Seiji Zenitani}
\affiliation{Space Research Institute, Austrian Academy of Sciences, Graz 8042, Austria}
\affiliation{Research Center for Urban Safety and Security, Kobe University, 1-1 Rokkodai-cho, Nada-ku, Kobe 657-8501, Japan}
\author{Shuichi Matsukiyo}
\affiliation{Faculty of Engineering Sciences, Kyushu University 6-1 Kasuga-Koen, Kasuga, Fukuoka 816-8580, Japan}
\affiliation{International Research Center for Space and Planetary Environmental Science, Kyushu University, Motooka, Nishi-Ku, Fukuoka 819-0395, Japan}
\author{Mami Machida}
\affiliation{Department of Astro-fusion Plasma Physics (AFP), Headquarters for Co-Creation Strategy, National Institute of Natural Sciences, Tokyo 105-0001, Japan}
\affiliation{Division of Science, National Astronomical Observatory of Japan, 2-21-1 Osawa, Mitaka, Tokyo 181-8588, Japan}
\author{Kazuhiro Sekiguchi}
\affiliation{Department of Astro-fusion Plasma Physics (AFP), Headquarters for Co-Creation Strategy, National Institute of Natural Sciences, Tokyo 105-0001, Japan}
\author{Amitava Bhattacharjee}
\affiliation{Department of Astrophysical Sciences, Princeton University, Princeton, New Jersey 08544, USA}
\affiliation{Department of Astro-fusion Plasma Physics (AFP), Headquarters for Co-Creation Strategy, National Institute of Natural Sciences, Tokyo 105-0001, Japan}
\affiliation{Princeton Center for Heliophysics, Princeton, New Jersey 08543, USA}
\affiliation{Princeton Plasma Physics Laboratory, Princeton University, Princeton, New Jersey 08540, USA}

\date{\today}

\begin{abstract}

Magnetic reconnection is an important source of energetic particles in 
systems ranging from astrophysics to the laboratory. The large separation of 
spatiotemporal scales involved makes it critical to determine the minimum 
physical model containing the necessary physics for modeling particle 
acceleration. By resolving the energy 
gain from ideal and nonideal magnetohydrodynamic electric fields
self-consistently in kinetic particle-in-cell
simulations of reconnection, we conclusively show the dominant 
role of the nonideal field for the early stage of energization known as 
injection. The importance of the nonideal field increases with 
magnetization, guide field, and in three-dimensions, indicating its general
importance for reconnection in natural astrophysical systems.
We obtain the statistical properties of the injection process from the
simulations, paving the way for the development of extended MHD models 
capable of accurately modeling particle acceleration in large-scale systems.
The novel analysis method developed in this study can be applied broadly
to give new insight into a wide range of processes in plasma physics.

\end{abstract}


%\maketitle

\section{Introduction}

Magnetic reconnection is known to be an important mechanism for converting 
magnetic field energy into plasma heating and nonthermal energetic particles
in space physics and astrophysics \cite{Zweibel2009,Yamada2010,Ji2011}.
For the relativistic magnetic field strengths of highly
magnetized astrophysical objects, reconnection has been
shown to be an efficient mechanism for producing power law
distributions of energetic particles through the use of
fully kinetic particle-in-cell (PIC) simulations
\cite{Zenitani2001,Sironi2014a,Guo2014}.
However, due to computational expense, 
fully kinetic simulations can only model small system sizes and 
short timescales, and it is a major challenge to relate such simulations to 
the global scales of realistic objects. An important question for studying 
particle acceleration from reconnection is whether fully kinetic simulations 
are necessary, or whether the important acceleration mechanisms can be 
captured in reduced models such as magnetohydrodynamics (MHD).
The answer to this question has deep implications for the modeling of
astrophysical objects, as only MHD models (in their traditional or extended forms)  can feasibly capture the global 
scales of realistic systems. Nonideal 
electric fields, defined as those that are not captured in ideal MHD,
are strongest inside the current sheet due to kinetic effects. Recent studies
have come to differing conclusions on the importance of these fields in the
relativistic regime, with arguments that these fields can be neglected \cite{Guo2019,Guo2022}, or in
contrast, that they are crucial for giving particles their initial boost to 
relativistic energies which injects them into Fermi acceleration processes \cite{Sironi2022}.
The majority of the overall particle energy gain is believed to come
from Fermi acceleration by ideal electric fields associated with the merging and contraction of plasmoids that are captured in ideal MHD \cite{Drake2006b}, however, this has
not been systematically tested. Particles continue to be exposed to
nonideal fields at all stages of their evolution, for example in secondary current sheets that form between merging plasmoids
\cite{Oka2010}, polarization electric fields from charge separation,
or from the rich variety of 
kinetic plasma waves that can be excited in the current sheet.

Past work analyzing the role of nonideal fields in magnetic 
reconnection has almost exclusively relied on approximate criteria, by
focusing on regions where the electric field magnitude exceeds the 
magnetic field magnitude ($\left |\bf{E} \right | > \left | \bf{B} \right |$) for antiparallel 
reconnection, or regions where the electric field has a component parallel to the magnetic 
field (${\bf E} \cdot {\bf B} \neq 0$) for reconnection with a finite guide 
field. While the ideal MHD condition ${\bf E} = - {\bf v}_{fl} \times {\bf B} / c$ (where ${\bf v}_{fl}$ is the plasma fluid velocity) is
manifestly violated in these two cases, these sufficient but not 
necessary conditions do not capture the full nonideal
field which is given exactly by ${\bf E} + {\bf v}_{fl} \times {\bf B} / c$.
The exact nonideal field can be obtained in postprocessing at a given
timestep from standard kinetic PIC simulation outputs, 
however, accessing the nonideal field during the runtime of a PIC simulation
requires a significant modification of the code with many subtleties
that must be considered. Furthermore, the necessity of calculating
an additional four quantities from the simulation particles (three fluid 
velocity components and the particle number density for each species)
to obtain the nonideal field adds a significant computational burden
to an electromagnetic PIC 
simulation, where three current density components alone are sufficient to advance the electromagnetic 
fields to the next timestep. For these reasons, approximate conditions
have remained the standard for investigating nonideal effects.

However, Figure \ref{fig:overview} (a) shows why it is critical to calculate the 
exact nonideal electric field for the purpose of investigating the limitations
of MHD. This figure shows the separation of the ideal (top) and nonideal (middle) electric
fields in the out-of-plane direction along with the binary result of the
$\left |\bf{E} \right | > \left | \bf{B} \right |$ condition (bottom)
calculated from a standard relativistic reconnection PIC simulation, zoomed 
into the diffusion region. The nonideal component has a characteristic 
magnitude that is typicaly around 5 times smaller than that of the ideal 
component, and shows an intricate structure extending throughout the current
sheet and into the reconnection outflows. Importantly, as seen in the bottom
plot of
Figure \ref{fig:overview} (a), there are significant 
contributions even in regions where the magnitude of the electric field does
not exceed that of the magnetic field, demonstrating the severe limitation
of the simplified $\left |\bf{E} \right | > \left | \bf{B} \right |$ 
condition. The nonideal field around X-points has been extensively studied \cite{Hesse2011TheReconnection}, however the physical mechanisms underlying the nonideal fields that are seen further downstream are not yet well understood.

In this Letter, we present the first analysis resolving energization from 
ideal and nonideal fields in magnetic reconnection self-consistently for 
every particle at every timestep in a PIC simulation. By determining the 
relative roles of the ideal and nonideal fields in the early stages of 
particle energization, we conclusively show the importance of the nonideal 
field in the injection process and determine a characteristic injection 
energy. By analyzing a set of simulations with various parameters of
astrophysical relevance, we find the importance of the nonideal field 
increases with magnetization, guide field strength, and with the 
inclusion of three-dimensional effects, indicating its general relevance for
astrophysical systems in nature. Numerical convergence studies show the 
robustness of the results and validity of the analysis
method, which can now be applied to give new insight into a wide range of 
scenarios in plasma physics. Finally, we obtain statistical
properties of the injection process, and discuss prospects for embedding
them into extended MHD simulations which could enable the modeling of 
astrophysical particle acceleration in large-scale systems.

\section{Methods}
\subsection{Simulation setup}

To model relativistic reconnection we use OSIRIS, a state-of-the-art, fully
relativistic and electromagnetic PIC code \cite{Fonseca2002a,Fonseca2008,Fonseca2013}. The simulations are initialized as
force-free current sheets \cite{Kilian2020}, with magnetization ranging from 
$\sigma = 12.5 - 200$ and guide field ranging from $B_{G} / B_{0} = 0-1$,
where $\sigma = B_{0}^{2} / 4 \pi n m c^{2}$ and $B_{0}$ is the asymptotic value of the initial reconnecting field.
The plasma consists of electron-positron pairs that are initially 
relativistically cold ($u_{th} / c = 10^{-4}$), and reconnection starts
from thermal noise in the system with no imposed perturbation.
(Unlike the case of an ionic plasma, the characteristic spatial scales of oppositely charged particles are identical in a pair plasma.) 
The initial current sheet half-widths are $\delta / (c/\omega_{p}) = 5$
for simulations with $\sigma = 12.5-50$, and
$\delta / (c/\omega_{p}) = 7, 10$ for simulations with $\sigma = 100, 200$,
respectively. Particles with initial positions
within $\pm 4 \delta$ from the center of the layer are excluded
from all analysis to eliminate effects from the initial current supporting
the magnetic field reversal. Unless otherwise stated, for two-dimensional (2D)
simulations the number of 
particles-per-cell is $n_{ppc} = 16$ per-species, the spatial and 
temporal resolutions are $\Delta x / (c/\omega_{p})= 0.25$ and
$\omega_{p} \Delta t = 0.175$, and the domain size is
$L_{x} / (c / \omega_{p}) = L_{y} / (c / \omega_{p}) = 1000$, where $x$ and
$y$ are the reconnection outflow and inflow directions, respectively. For
three-dimensional (3D) simulations, $n_{ppc} = 8$ per-species, 
$\Delta x/(c/\omega_{p}) = 0.25$,
$\omega_{p} \Delta t = 0.143$, $L_{x} / (c / \omega_{p}) = L_{y} / (c / \omega_{p}) = 1000$, 
and $L_{z} / (c / \omega_{p}) = 150$. The $y$ boundaries are conducting for
fields and reflecting for particles, and the $x$ and $z$ boundaries are 
periodic.
Cubic particle shapes are used to interpolate between particle and grid 
quantities for all simulations. In the following analysis, particle
energies are normalized to $mc^{2}$.

\subsection{A new analysis method}

To analyze the nonideal electric field, we have modified OSIRIS to separate
the ideal and nonideal components of the electric field at every timestep
in the simulation.  First, the plasma fluid velocity is calculated as the
velocity of the Eckart frame \cite{Zenitani2018} (where the particle flux vanishes) relative to the lab frame, on a discrete grid from the 
positions and time-centered velocities of all of the particles in the
simulation as 
${\bf v}_{fl} \left ( {\bf X}_{i} \right )=  \left ( \sum_{e^{-},e^{+}}\sum_{n=1}^{N_{p}} S \left ( {\bf x}_{n} - {\bf X}_{i} \right ) m {\bf u}_{n} / \gamma_{n} \right ) / \left ( \sum_{e^{-},e^{+}}\sum_{n=1}^{N_{p}}  S \left ( {\bf x}_{n} - {\bf X}_{i} \right ) m \right )$, where
$S$ is the particle shape factor \cite{Birdsall}.  
Due to the time staggering characteristic of the leapfrog
particle pusher used in PIC simulations, it is necessary to retain the 
particle velocity from the previous timestep to calculate the time-centered 
velocity.
The fluid velocity and magnetic field are then interpolated to the position
of the particle, allowing a decomposition of the electric field acting
on the particle into ideal and nonideal components as 
${\bf E}_{\mathrm{I}} = - {\bf v}_{fl} \times {\bf B} / c$ and
${\bf E}_{\mathrm{N}} = {\bf E}_{\mathrm{T}} - {\bf E}_{\mathrm{I}}$,
respectively.
The interpolation between particle and grid quantities used at all steps is 
the same as that in the simulation (cubic). With access to the ideal and 
nonideal components of the electric field, their contributions to the energy
gain of a particle in a given timestep can be calculated as
$\Delta \epsilon_{N} = q {\bf E}_{\mathrm{N}} \cdot {\bf v}_{\mathrm{particle}} \Delta t$, and
similarly for the ideal component. Finally, a normalization factor is applied
to ensure the sum of the energy gain from the ideal and nonideal components is
exactly equal to the total energy gain of the particle between timesteps,
however we have found the results are not sensitive to the small 
discretization error introduced without normalization. We 
track the cumulative ideal and nonideal energization for every particle at 
every timestep in the simulation to study their relative importance in 
different acceleration processes.

\subsection{Testing injection} \label{ssec:injection_analysis}

\begin{figure}[htp]
\begin{center}
\includegraphics[width=0.95\textwidth]{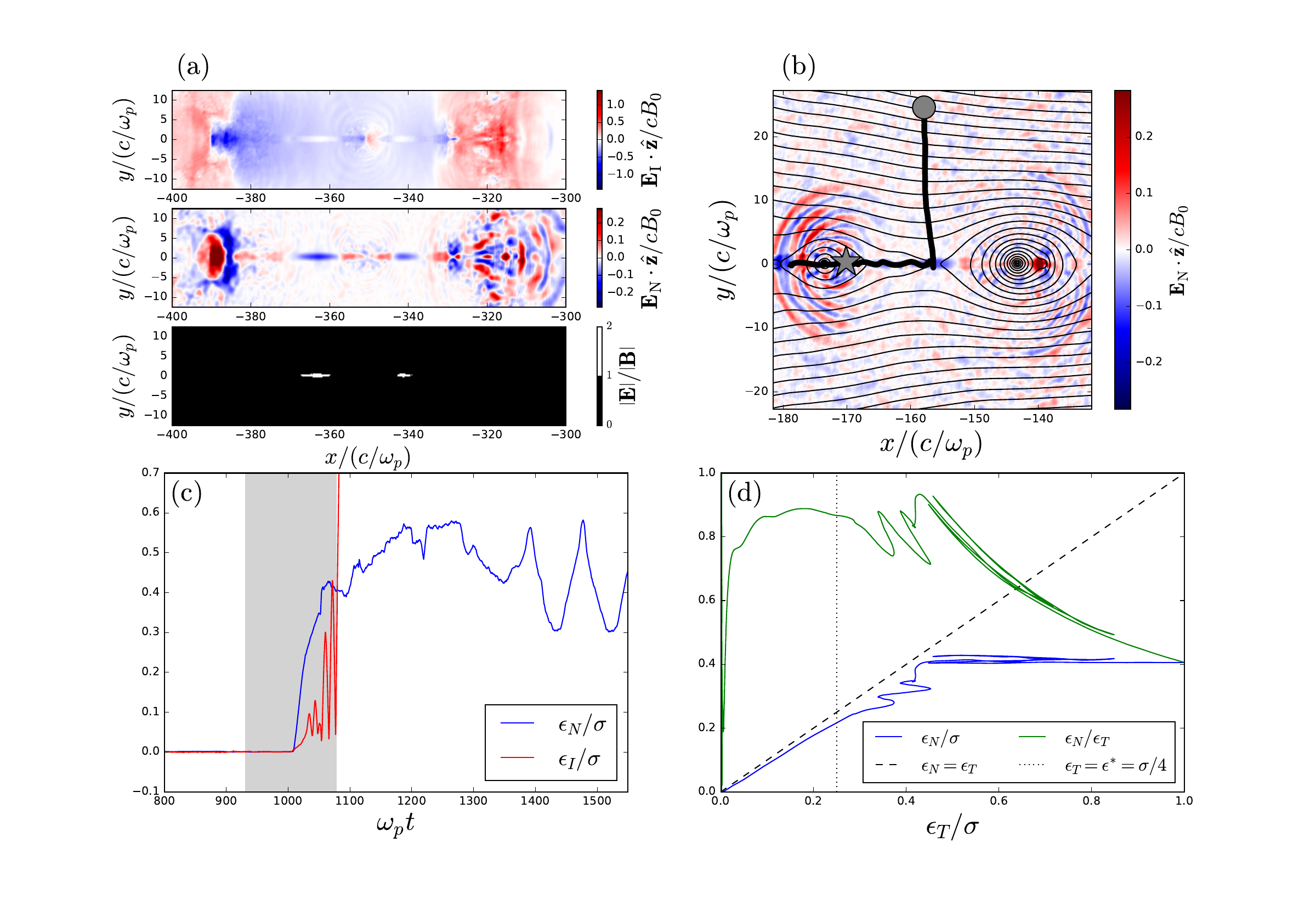}
\caption{\label{fig:overview} 
(a) Physical quantities from a region of 
a $\sigma = 50$ relativistic reconnection simulation: ideal electric field 
${\bf E}_{\mathrm{I}}$ (top); nonideal electric
field ${\bf E}_{\mathrm{N}}$ (middle); ratio of electric
and magnetic field magnitudes with color scaled so that only regions satisfying
$\left |\bf{E} \right | > \left | \bf{B} \right |$ are white
(bottom).
(b) Representative simulation
particle trajectory from a $\sigma = 50$ simulation of relativistic 
reconnection plotted over the out-of-plane component of the
nonideal electric field with overlayed magnetic field lines.
The trajectory corresponds to the shaded times in 
(c), with the circle showing the position at the beginning of this period
and the star showing the position at the time when $\epsilon_{T}=\sigma/4$.
(c) Evolution of energy gained by the nonideal
($\epsilon_{N}$) and ideal ($\epsilon_{I}$) electric
fields. (d) Energy gained from the nonideal field ($\epsilon_{N} / \sigma$) versus total energy ($\epsilon_{T} / \sigma$) and fractional energy gain from the nonideal
electric field, $\epsilon_{N}/ \epsilon_{T}$, vs. total particle energy $\epsilon_{T} / \sigma$.}
\end{center}
\end{figure}

To study the critical problem of injection, we isolate the contribution
of the nonideal field in the early stages of particle energization.
Figure \ref{fig:overview} shows an example trajectory of a representative
simulation particle plotted over the out-of-plane component of the nonideal
electric field in (b),
along with the energy gained from the 
nonideal ($\epsilon_{N}(t) = \sum_{0}^{t} \Delta \epsilon_{N}$) and
ideal ($\epsilon_{I}(t) = \epsilon_{T}(t)-\epsilon_{T}(0)-\epsilon_{N}(t)$, where $\epsilon_{T}(t)$ is the particle's total energy with initial energy
$\epsilon_{T}(0) \ll \sigma$) electric fields as a function of time in (c).  The particle enters the current sheet and first becomes 
strongly energized by the nonideal component of the electric field.  After 
reaching an energy of $\approx 0.4 \, \sigma$, the ideal component of the 
electric field takes over as the dominate source of energization. Figure \ref{fig:overview} (d) shows a plot of $\epsilon_{N}/\sigma$
versus the total particle energy $\epsilon_{T}/\sigma$ in blue.  During the initial stage,
$\epsilon_{N} \approx \epsilon_{T}$, indicating the acceleration is dominantly
nonideal.  To determine the importance of the nonideal field in the early 
stages of energization we first define a threshold energy $\epsilon^{*}$.
The fraction of energy gained from the nonideal field when the particle 
reaches the threshold energy, $(\epsilon_{N} / \epsilon_{T})|_{\epsilon_{T}=\epsilon^{*}}$,
gives a quantified measure of the importance of the nonideal field for 
acceleration below particle energies $\epsilon^{*}$.  Figure 
\ref{fig:overview} (d) shows
the calculation of this quantity for the particle trajectory under
consideration with a threshold set to be $\epsilon^{*} = \sigma / 4$.
In this case $(\epsilon_{N} / \epsilon_{T})|_{\epsilon_{T}=\epsilon^{*}} \approx 0.8$,
indicating the nonideal field was primarily responsible for bringing
the particle to the energy of $\sigma / 4$. Calculating
$(\epsilon_{N} / \epsilon_{T})|_{\epsilon_{T}=\epsilon^{*}}$ for all particles in a simulation
and for various values of $\epsilon^{*}$ allows a statistical determination
of the role of the nonideal field in various energy ranges, and in particular
its role during particle injection when $\epsilon^{*}$ is chosen to be
suitably small.

\section{Results}

\begin{figure}[htp]
\begin{center}
\includegraphics[width=0.95\textwidth]{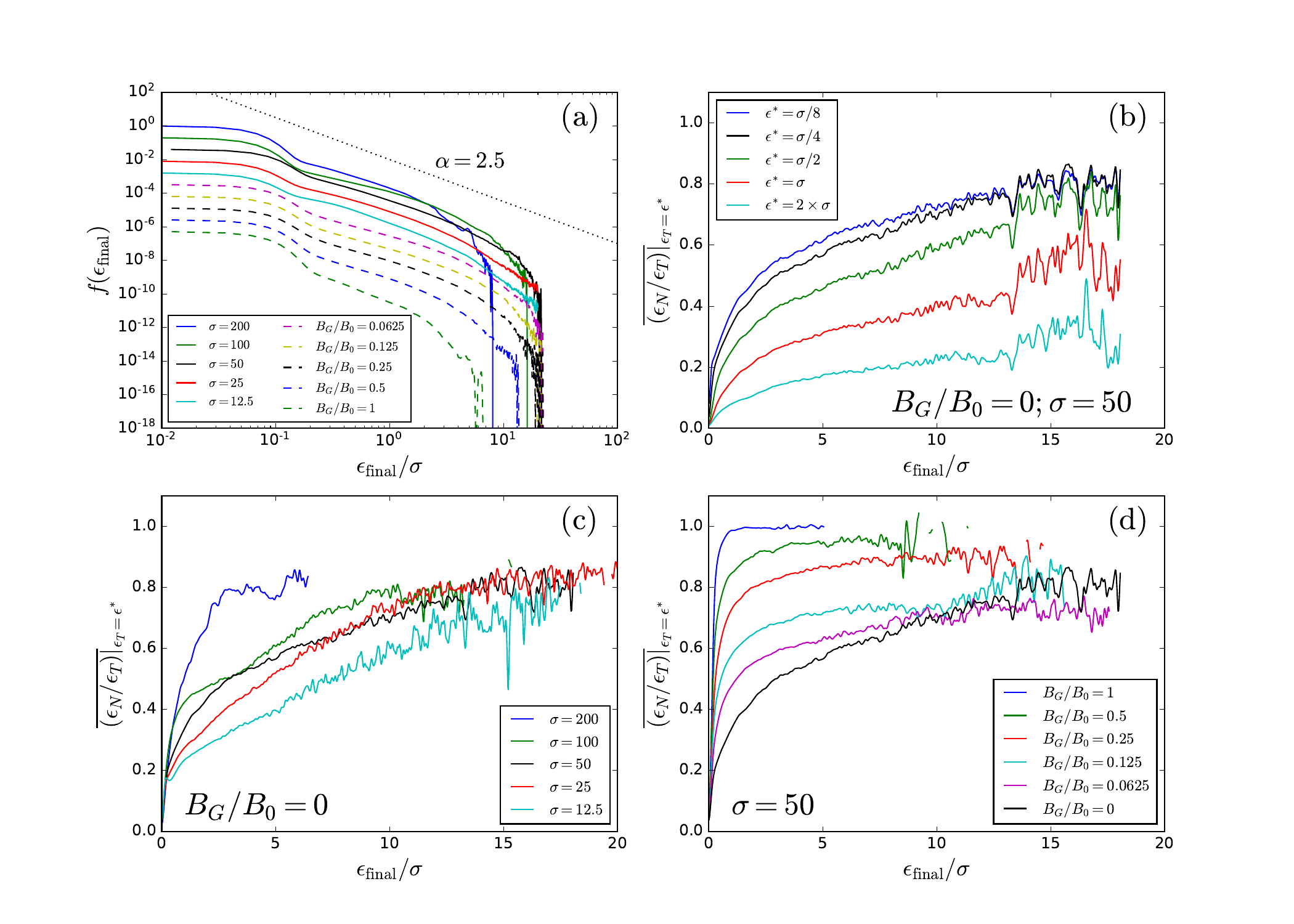}
\caption{\label{fig:scans} (a) Electron energy spectra at the end of the simulation
($\omega_{p} t \approx 2000$) for various conditions. Solid lines have $B_{G}=0$, and dashed
lines have $\sigma = 50$ and $B_{G} >0$. Dotted black line shows
a power law spectrum with index $\alpha=2.5$ for reference.
(b) Average fractional nonideal contribution to 
energization up to various threshold energies $\epsilon^{*}$ for the $\sigma = 50$,
$B_{G} = 0$ simulation. (c) Average fractional nonideal contribution to 
energization up to $\epsilon^{*}=\sigma/4$ for simulations with varied $\sigma$ and
$B_{G} = 0$. (d) Same as (c), but for $\sigma=50$ and varied $B_{G}$.
All solid black lines
correspond to the 2D simulation with $\sigma=50$ and $B_{G}=0$.
}
\end{center}
\end{figure}

We now apply this analysis
to a set of 2D simulations with varied magnetization $\sigma$
and guide field $B_{G}$.  All analysis is applied at the end of the
simulations at $\omega_{p} t \approx 2000$. The
corresponding electron energy spectra are shown in Figure \ref{fig:scans} (a),
with normalizations shifted for ease of comparison.
All spectra show nonthermal components resembling power law distributions
$\sim \epsilon_{\mathrm{final}}^{-\alpha}$ over more than a decade in energy normalized to
sigma.
The cases with $B_{G} = 0$ have power law indices
in the range $\alpha = 2-3$ and maximum particle energies in the range
$\epsilon_{max}/\sigma\approx 8-20$.
The simulations with finite guide field all have $\sigma = 50$, and their
power laws become steeper and extend to lower maximum energies as the guide
field is increased. The spectrum for the $\sigma = 200$ is slightly less developed due to the wider current sheet width used which increases the time for reconnection onset. We first focus on the simulation with $\sigma = 50$
and $B_{G} = 0$.  The curves in Figure \ref{fig:scans} (b) show the
fractional contribution of the nonideal field to the energy gain up to the
threshold energy, averaged over all electrons and binned by final particle 
energy $\epsilon_{\mathrm{final}}$:
$\overline{(\epsilon_{N} / \epsilon_{T})|}_{\epsilon_{T}=\epsilon^{*}} = (1/N_{p})\sum_{i=1}^{N_{p}} (\epsilon_{N,i} / \epsilon_{T,i})|_{\epsilon_{T,i}=\epsilon^{*}}$.
For particles with a given final particle energy
$\epsilon_{\mathrm{final}}$, the corresponding
points on the curves quantify the role of the
nonideal field in bringing those particles to
the energy $\epsilon^{\star}$.
Each curve in this panel is obtained from all of the electron
data at a single time in the same simulation, but using different values
of the threshold energy in the range $\epsilon^{*}/\sigma = 1/8 - 2$.
For the smallest threshold energy of $\epsilon^{*}/\sigma = 1 / 8$ (blue 
curve) the nonideal field begins to dominate for particles with final 
energies $\epsilon_{\mathrm{final}} \gtrsim 2\sigma$, and increases in importance with final
particle energy.  For the particles with the highest energy, the nonideal
field provided more than $80\%$ of the energy in their initial rise to
$\epsilon_{T} = \sigma / 8$.  As the threshold energy is increased the fraction
of energy gained by the
nonideal field begins to drop. This indicates the ideal field is primarily
responsible for later stages of energization, in accordance with previous
expectations. The fraction of energy coming from the nonideal field steeply drops for
threshold energies of $\epsilon^{*} > \sigma /4$, which gives evidence of a critical injection energy of $\epsilon^{*} \approx \sigma / 4$ that allows particles to reach large energies.  This trend is confirmed for
simulations of $\sigma = 25$ and $\sigma = 100$ in addition to the $\sigma=50$
case presented in Figure \ref{fig:scans} (b).
For all of the following analysis, we fix the threshold energy to 
$\epsilon^{*}  = \sigma / 4$ and refer to
energization up to level as injection.

\begin{figure}[htp]
\begin{center}
\includegraphics[width=0.95\textwidth]{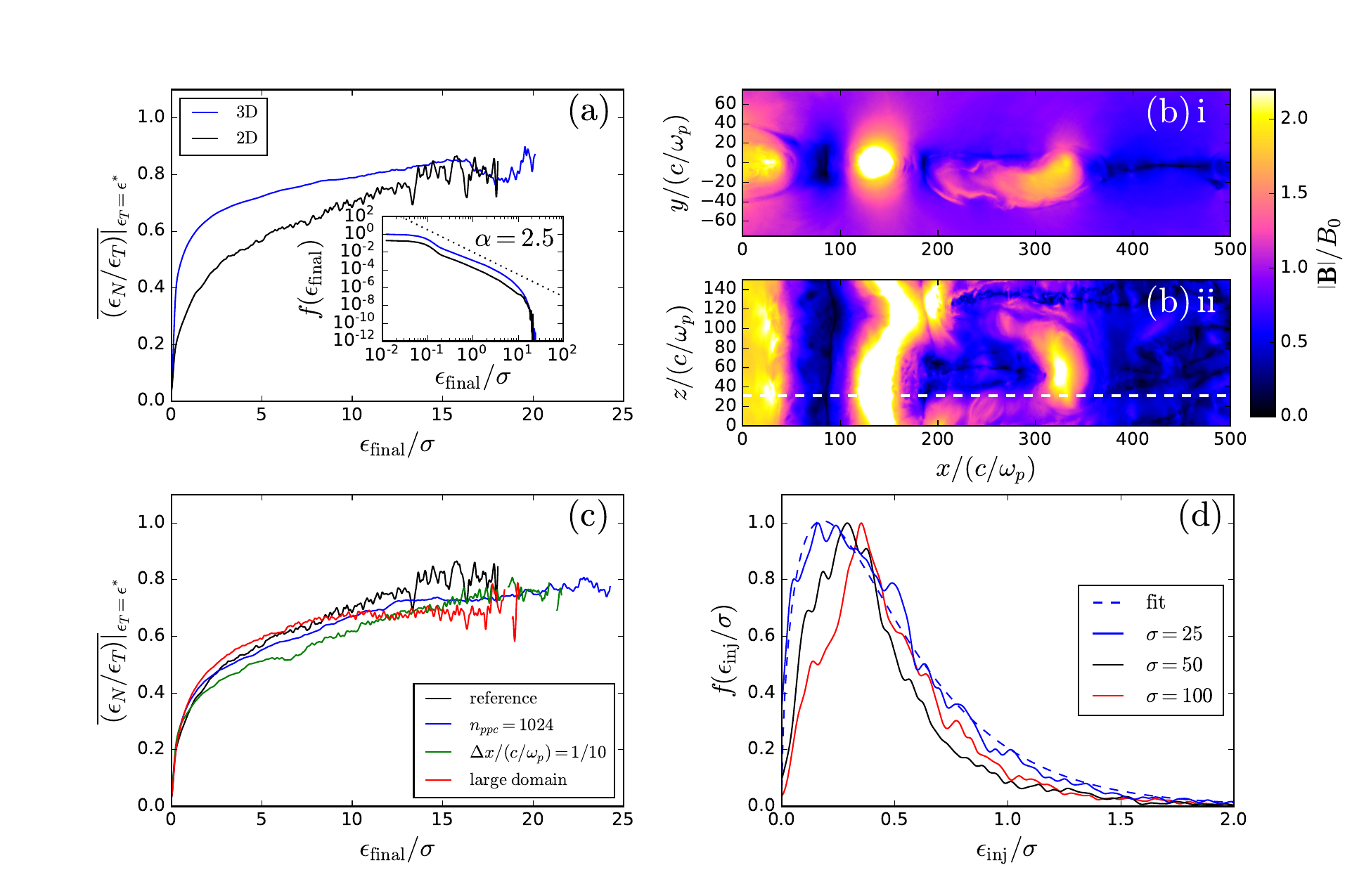}
\caption{\label{fig:3d_convergence} (a) Comparison of average fractional contribution of
nonideal field to injection for 2D and 3D simulations with $\sigma=50$ and $B_{G}=0$, at
time $\omega_{p} t \approx 2000$. Inset shows the corresponding electron energy spectra, with a
power law spectrum with index $\alpha=2.5$ plotted for reference.
(b) Magnetic field magnitude in 2D slices of the 3D simulation at 
$\omega_{p} t \approx 1200$. (i) $x-y$ plane at $z/(c/\omega_{p})=31$,
and (ii) $x-z$ plane at $y=0$ (dashed line shows the location of the
$x-y$ slice). (c) Convergence studies showing the 
average fractional contribution of the nonideal field to injection for
$\sigma = 50$, $B_{G} = 0$ simulations with standard parameters (black),
$n_{ppc}=1024$ (blue), $\Delta x/(c/\omega_{p})=1/10$ (green), and a domain 
size of $L_{x}/ (c/\omega_{p}) = 2000$ and $L_{y}/ (c/\omega_{p}) = 4500$
(red). (d) Distribution of injection energies obtained from 
simulations with varied sigma.  Dashed blue line shows a fit of the $\sigma=25$ 
case to the function $a\epsilon^{b} e^{-\epsilon/c}$. All solid black lines
correspond to the 2D simulation with $\sigma=50$ and $B_{G}=0$.}
\end{center}
\end{figure}

Next we analyze five different simulations with magnetization ranging
from $\sigma = 12.5 - 200$ and zero guide field, with each curve in
Figure \ref{fig:scans} (c) corresponding to one of these simulations.
The nonideal field invariably dominates the injection for the highest energy
particles, and shows a trend of increasing importance for higher $\sigma$.
A similar comparison is shown in Figure \ref{fig:scans} (d), now for six
simulations with $\sigma = 50$ and guide field ranging from
$B_{G}/B_{0} = 0 - 1$.
Again, the nonideal field dominates the injection for the highest energy
particles in every simulation.  For larger guide fields the
importance of the nonideal field in injection is seen to sharply increase, 
with the nonideal field providing nearly $100 \%$ of the energy during injection
for particles that reach final energies of $\epsilon_{\mathrm{final}} > \sigma$ for
the $B_{G} / B_{0} = 1$ simulation. The
simulations considered thus far have all been 2D, however
the current sheet structure and dominant acceleration mechanism
are significantly modified in 3D \cite{Zhang2021}.  The presence of additional kinetic
instabilities in 3D which lead to the distortion of flux ropes and turbulence may also be
expected to enhance magnetic field diffusion in the current sheet and 
thereby strengthen nonideal effects. To understand the influence of 3D effects,
we next analyze a 3D simulation with $\sigma = 50$ and $B_{G} = 0$, shown
in Figures \ref{fig:3d_convergence} (a,b). The 3D simulation has the same size as 
the 2D simulations in the reconnection plane ($x-y$) that is modeled in 2D, and the
resulting electron spectrum shows a similar shape and maximum energy to the
analogous 2D case with $\sigma = 50$ and $B_{G} = 0$ at the end of the 3D simulation (Figure \ref{fig:3d_convergence} (a) inset).
In the additional third dimension the size is $L_{z} / (c / \omega_{p}) = 150$, which
is sufficient to capture significant distortion of the flux ropes and disruption of the
current sheet, seen in Figure \ref{fig:3d_convergence} (b). The fraction of injection energy
coming from the nonideal field is compared for 2D and 3D in Figure \ref{fig:3d_convergence}
(a), showing how the nonideal field indeed increases in importance in 3D,
providing nearly $90\%$ of the injection energy for the highest energy particles
and enhanced by $10-20\%$ over the 2D case at lower energies.

The method of obtaining the nonideal field used in this study relies on
calculating fluid quantities by averaging a finite number of simulation particles
onto a spatial grid, and thus it is critical to perform convergence
studies to ensure the results are not being affected by discrete particle 
noise or spatial resolution. Figure \ref{fig:3d_convergence} (c) shows a
convergence study of the nonideal injection analysis for 2D simulations with
$\sigma = 50$ and $B_{G} = 0$. When compared to the reference case 
analyzed previously (black line), the results do not significantly change when 
increasing the number of particles-per-cell per-species from $n_{ppc}=16$ to 
$n_{ppc}=1024$, or increasing the spatial resolution from
$\Delta x/(c/\omega_{p}) = 1/4$ to $\Delta x/(c/\omega_{p}) = 1/10$.  Finally, it is important
to understand if the system size or boundary conditions are influencing
the results. To test this we have run a simulation with a domain size
of $L_{x}/ (c/\omega_{p}) = 2000$ and $L_{y}/ (c/\omega_{p}) = 4500$, which
is sufficient to ensure light crosses the periodic $x$ boundaries exactly once over
the duration of the simulation and that light traveling from the $y$ boundaries first reaches the edge of
the largest plasmoids only at the end of the simulation.
The insensitivity of the results to this change (red
curve in Figure \ref{fig:3d_convergence} (c)) confirms that the
boundaries are not influencing the results.

\section{Discussion}

The analysis presented above definitively demonstrates the critical role 
of the nonideal electric field in particle injection for relativistic
reconnection, providing the dominate source of
energy during injection for the wide range of parameters studied. The 
increase in the importance of the nonideal field with $B_{G}$ and in 3D
indicates nonideal fields will be responsible for injecting particles in
the general physical scenario of arbitrarily aligned magnetic fields in 
a 3D plasma.  Although nonideal fields are by definition beyond those captured in MHD, it may be possible to develop test particle models that
approximate the nonideal effects of particle injection and allow the
remaining acceleration to come from the ideal fields of MHD. A key component of such a model would be the
distribution of injection energies, which can be obtained from the 
trajectories of PIC simulation particles as the amount of energy gained 
from the nonideal field at the time when injection ceases and the
ideal field takes over as the dominant source of energy.  
An approximate criteria for determining the end of injection
is the time when $d \epsilon_{N} / d \epsilon_{T} = 0$, which can be calculated for
each particle during the simulation to obtain a distribution of
injection energies.
For the 
example particle trajectory in Figure \ref{fig:overview}, the injection energy can be seen to be $\epsilon_{\mathrm{inj}}/\sigma\approx 0.4$ 
from panel (d). 
Figure \ref{fig:3d_convergence} (d) shows the 
distribution of injection energies obtained in this way from the
electrons for three simulations with varied $\sigma$. 
With increasing $\sigma$ the peak energy of the distribution shows a 
modest increase, accompanied by a decrease in the peak width. All peaks
are near $\epsilon_{\mathrm{inj}} / \sigma \approx 1/4$, consistent with the
previous determination from Figure \ref{fig:scans} (b) that the
critical injection energy is $\epsilon_{\mathrm{inj}} / \sigma \approx 1/4$.
The distribution can be approximated by a gamma distribution
parameterized as
$a\epsilon^{b} e^{-\epsilon/c}$, and the dashed line shows the 
result of a least squares fit to the $\sigma = 25$ case which yielded
$a = 4.68$, $b=0.566$, and $c=0.320$.  A full characterization of
the distribution of injection energies, its dependence on the plasma 
conditions, and the incorporation into MHD test particle simulations
will be the subject of future work. 
The present analysis does not distinguish between different sources of nonideal fields, however the same technique could be applied to precisely determine the roles of each term in a generalized Ohm's in different stages of particle acceleration.  This would be particularly valuable for connecting to high-moment fluid models \cite{Hakim2008ExtendedEquations,Dong2019,Ng2020AnInstabilities,Wang2018ElectronMagnetosphere}.
Finally, we note that the convergence
studies demonstrate the soundness and computational feasibility of this
novel analysis method, which can now be applied more broadly to other particle acceleration
scenarios such as collisionless shock waves, turbulence, and relativistic
jets.

\section{Conclusions}

In conclusion, we have performed the first self-consistent analysis
of the energization from ideal and nonideal electric fields from
relativistic reconnection within fully kinetic PIC simulations.
We have done so by rigorously separating nonideal and ideal electric fields from first principles without adopting ad hoc definitions of the nonideal field. We
find the nonideal field is the dominate source of energy for particles
during the injection process, with later energization coming from the
ideal field. The importance of the nonideal field for injection increases
with $\sigma$, $B_{G}$, and in 3D, indicating the general relevance of
these results for relativistic reconnection in nature. We obtain
the statistical distribution of injection energies from the simulation
particles, which may be incorporated into extended test particle MHD
models that allow accurate simulation of particle acceleration in
large-scale astrophysical systems. Convergence studies demonstrate the
validity of this novel analysis method, which may now be applied to yield new 
insight into a wide variety of plasma processes.

%\acknowledgments{}
\section{Acknowledgments}
This project was supported by the International Research Collaboration Center,
Astro-fusion Plasma Physics (IRCC-AFP) program of the National
Institutes of Natural Sciences (NINS), Japan. S.T. and A.B. acknowledge gratefully the support of NSF Awards 2206756 and 2209471.
A part of this research was supported by JSPS KAKENHI (S.Z.: 21K03627; S.M.: 22H01287; M.M.: 19K03916, 20H01941, 22H01272).
Simulations were performed on
Perlmutter (NERSC). The authors acknowledge the OSIRIS Consortium, consisting
of UCLA and IST (Portugal) for the use of the OSIRIS 4.0 framework.
We thank R. Matsumoto, K. Tomida, and M. Hoshino for valuable discussions.

\bibliography{nonideal_acceleration}

\end{document}